\documentclass[twoside]{dis09}
\usepackage[latin1]{inputenc}
\usepackage[dvips]{graphicx,epsfig,color}
\usepackage{wrapfig,rotating}
\usepackage{amssymb,amsmath,array}
\usepackage{booktabs}
\usepackage{mcite}

\begin{document}
\title{Extraction of $\mathbf{F_{2}^{c}(x,Q^2)}$ from $\mathbf{D^{\mathbf{*\pm}}}$ cross sections at H1}

%***********************************************************************
% AUTHORS INFORMATION AREA
%***********************************************************************
\author{Andreas Werner Jung% and Second Author$^2$
\thanks{Supported by the German Federal Ministry of Science and
Technology under grant 05H16VHB.} ~for the H1 Collaboration
%
% DO NOT MODIFY THE FOLLOWING '\vspace' ARGUMENT
\vspace{.3cm}\\
%
% Addresses and institutions (remove "1- " in case of a single institution)
University of Heidelberg - Kirchhoff-Institute for Physics \\
Im Neuenheimer Feld 227, 69120 Heidelberg - Germany
%
% Remove the next three lines in case of a single institution
% \vspace{.1cm}\\
% 2- University of Heidelberg - Kirchhoff-Institute for Physics \\
% Im Neuenheimer Feld 227, 69120 Heidelberg - Germany
}
%***********************************************************************
% END OF AUTHORS INFORMATION AREA
%***********************************************************************

\maketitle

\begin{abstract}
The charm contribution to the proton structure, $F_2^{c}\left(x,Q^2\right)$, is determined using the inclusive cross sections of $D^{*\pm}(2010)$ meson production in deep-inelastic scattering. The cross section measurement covers the region $5<Q^2<1000~\rm{GeV^2}$ in photon virtuality and $0.02<y<0.70$ in the inelasticity of the scattering process. The $D^{*\pm}$ meson is measured in transverse momentum and pseudo-rapidity down to $p_{_{T}} > \mbox{1.5}~\rm{GeV}$ and up to $|\eta| < \mbox{1.5}$. The data were taken with the H1 detector corresponding to an integrated luminosity of $347~\rm{pb^{-1}}$. $F_2^c$ is determined from the $D^*$ production cross sections and compared to leading and next-to-leading order perturbative QCD predictions. %The extrapolation uncertainties are studied varying the theory parameters like charm mass, renormalisation and factorisation scales as well as the fragmentation model. 
\end{abstract}

\section{Introduction}
The charm quark production in electron-proton ($ep$) scattering is dominated by the boson-gluon-fusion (BGF) process $(\gamma p \rightarrow c\bar{c})$. The charm contribution, $F^{c}_2(x,Q^2)$, to the proton structure is obtained by using the expression for the one photon exchange cross section for charm production:
\begin{equation}
\displaystyle
\frac{d^2\sigma^{c}}{dxdQ^2}=\frac{2\pi\alpha_{em}^2}{Q^4x}
\left(1+\left(1-y\right)^2\right)\;F^{c}_2(x,Q^2)~.
\end{equation}
The $y$ range of the measurements is such that the contribution from the longitudinal structure function $F_L^c$ is negligible. The current analysis \cite{dis09:url} uses data taken with the H1 detector \cite{dis09:h1} during the HERA-II running period at a centre-of-mass energy of $\sqrt{s}=318~\mathrm{GeV}$. Preliminary results \cite{dis09:jung,dis09:brinkmann} of the inclusive $D^*$ cross section measurements at medium and high $Q^2$ are used. Compared to early results \cite{dis09:2001zj} by H1 a larger phase space and a significantly larger integrated luminosity of $347~\mathrm{pb^{-1}}$ has been used. Hence, more precise tests of perturbative QCD (pQCD) become possible.

\section{Theoretical Models of Open Charm Production}
The description of heavy flavour production in $ep$ collisions is based on pQCD at leading order (LO) or at next-to-leading order (NLO) for which calculations in several schemes are available \cite{dis09:riemersma,*dis09:HVQDIS, dis09:harris, *dis09:acot, *dis09:collins}. 
%%% All approaches assume the scale to be hard enough to perform pQCD calculations and to guarantee the validity of the factorisation theorem. \\
%In leading order (LO) pQCD, the BGF process ($\gamma g\rightarrow Q\overline Q$) is the dominant contribution. This treatment has been extended to next-to-leading order (NLO) for which calculations in several schemes are available \cite{riemersma,riemersma2,harris,acot,collins}. 
In this analysis, the $D^{*\pm}$ meson cross sections in the visible range, as well as the charm contribution to the proton structure function are calculated using two different models. On the one hand the HVQDIS program \cite{dis09:riemersma,*dis09:HVQDIS} is used. It is based on a NLO calculation in the fixed-flavor-number-scheme (FFNS) providing differential cross sections of massive charm quarks including mass effects at the production threshold. The parton evolution is performed according to  the DGLAP evolution equations \cite{dis09:dglap1, *dis09:dglap2, *dis09:dglap3}. HVQDIS applies independent fragmentation. On the other hand the LO Monte Carlo program CASCADE \cite{dis09:CASCADE}, supplemented with parton showers, based on the CCFM evolution scheme is used. The CCFM evolution equations~\cite{dis09:ccfm1,*dis09:ccfm2,*dis09:ccfm3,*dis09:ccfm4} are expected to be more appropriate to describe the parton evolution at small $x$. The hadronisation of partons is performed using the Lund String model as implemented in PYTHIA~\cite{dis09:PYTHIA61}. In case of HVQDIS the proton parton density function (PDF) \mbox{MRST04FF3nlo} \cite{dis09:mrst04ff} is used, while for CASCADE the proton PDF parametrisation A0~\cite{dis09:a0} is used. The renormalisation and factorisation scale for both pQCD calculations has been set to $\mu_{r,f}^2 = 4m_c^2+ Q^2$.

% In the parton cascade, gluons are 
% emitted in an angular ordered manner to account for coherence effects. Due to 
% this angular ordering, the gluon distribution depends
% on the maximum allowed angle in addition to the momentum fraction $x$ and the 
% transverse momentum of the propagator gluon. The cross section is then 
% calculated according to the $k_t$-factorization theorem by convoluting the 
% unintegrated gluon density with the off-shell photon gluon fusion matrix 
% element with massive quarks for the hard scattering process.

\subsection{Fragmentation Model}
Differential inclusive cross sections of charmed mesons are calculated using the HQVDIS program after fragmenting the charm quarks into $D^{*\pm}$ mesons. %%% in the photon-proton centre of mass frame into $D^{*\pm}$ mesons. 
The Kartvelishvili fragmentation function \cite{dis09:kart_frag} is used, which is controlled by a single parameter $\alpha$. This is also used for CASCADE, where higher charm resonances as determined by ALEPH~\cite{dis09:aleph} are included. The charm fragmentation function was measured at H1 using inclusive $D^{*\pm}$ meson production associated with jet production, where a different behaviour of the fragmentation function close to the threshold of charm production and far above the threshold is reported \cite{dis09:2008tt}. This results in different values for $\alpha$ (see Table \ref{tab:ParVar}), which suggests a $\hat{s}$-dependent charm fragmentation function. $\hat{s}$ denotes the invariant mass of the produced $c\bar{c}$ pair. Since the fragmentation model influences the kinematic distributions of the $D^*$ meson an additional uncertainty from the fragmentation model is assigned.

% Close to the charm threshold corresponding to $<\hat{s}>$=32 GeV$^2$, the fragmentation parameter $\alpha=6.0$ and otherwise $\alpha=3.3$ is used. Average value of $\hat{s}$=32 GeV$^2$ corresponds to the cutoff value of $\hat{s}<$70 GeV$^2$ in the HVQDIS calculation. Since the interpolation of the parameter $\alpha$ in the HVQDIS program is technically not possible, the cross over value of $\hat{s}$ is varied by $\pm$20 GeV$^2$ around the cut value. In addition the Kartvelishvili fragmentation parameter $\alpha$ is varied by the uncertainty of the measurement: $\alpha=6.0+1.1-1.3$ at $\hat{s}<$70 GeV$^2$ and $\alpha=3.3\pm0.4$ at $\hat{s}>$70 GeV$^2$. $\alpha=8.5$ and otherwise $\alpha=4.3$ is used. fragmentation parameter $\alpha$ is varied by the uncertainty of the measurement: $\alpha=8.2+1.2-1.1$ at $\hat{s}<$70 GeV$^2$ and $\alpha=4.5+0.6-0.5$ at $\hat{s}>$70 GeV$^2$. The uncertainties vary with $Q^2$ and amount to about 5\%.

%---- etxraction of F2c
%
\section{$\mathbf{D^{*\pm}}$ Cross Section Data \& Extraction of $\mathbf{F_2^c(x,Q^2)}$}
The cross section measurement covers the kinematic region of $5<Q^2<1000~\rm{GeV^2}$ and $0.02<y<0.70$ in the inelasticity of the scattering process.
\begin{figure}[ht]
%     \begin{minipage}[c]{0.475\columnwidth}
   \centerline{\includegraphics[width=0.88\columnwidth]{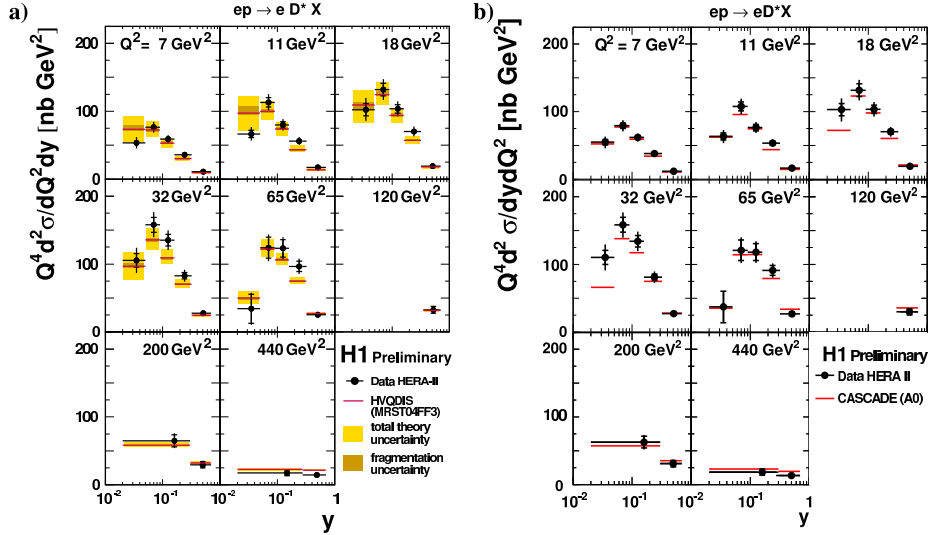}}
%      \end{minipage}
%      \hspace{.05\linewidth}
%      \begin{minipage}[c]{0.475\columnwidth}
%  \centerline{\includegraphics[width=1.0\columnwidth]{JungAndreas.fig2.eps}}
%      \end{minipage}
 \caption{\label{fig:crossSec_yQ2} The double differential cross section in $y$ and $Q^2$ compared to the prediction from HVQDIS a) and to the prediction provided by CASCADE b). The $\hat{s}$-dependent charm fragmentation function is used for both predictions. The shaded band in a) reflects the theoretical uncertainty estimated by parameter variations given in Table \ref{tab:ParVar}.}    
 \end{figure}
The visible range of the $D^{*\pm}$ meson measurement is restricted to $p_T (D^*) > \mbox{1.5}~\rm{GeV}$ and $|\eta (D^*)| < \mbox{1.5}$. The double differential cross section as a function of $Q^2$ and $y$ is shown in Figure \ref{fig:crossSec_yQ2}a) compared to the prediction from HVQDIS and in b) for CASCADE. Both describe the data reasonably well. A total experimental error of 9\% was achieved. These data are used for the extraction of $F_2^{c}$. The measured inclusive $D^{*\pm}$ cross sections $\sigma_{\rm{vis}}^{\rm{exp}}(y,Q^2)$ in bins of $y$ and $Q^2$ are converted to a bin centre corrected $F_2^{c~\rm{exp}}(\langle x\rangle,\langle Q^2\rangle)$ by the relation:
\begin{equation}
F_2^{c~\rm{exp}}(\langle x\rangle,\langle Q^2\rangle)=
\frac{\displaystyle \sigma_{\rm{vis}}^{\rm{exp}}(y,Q^2)}
{\displaystyle \sigma_{\rm{vis}}^{\rm{theo}}(y,Q^2)}\cdot
F_2^{c~\rm{theo}}(\langle x\rangle,\langle Q^2\rangle)~,
\label{eqn:f2cexp}
\end{equation}  
where $\sigma_{\rm{vis}}^{\rm{theo}}$ and $F_2^{c~\rm{theo}}$ are the theoretical predictions from the model under consideration. The Bj\o rken variable $x$ is related to $y$ via $x=Q^2 / y \cdot s$. The contribution of open beauty
\begin{wrapfigure}{r}{0.56\columnwidth}
 \centerline{\includegraphics[width=0.4\columnwidth]{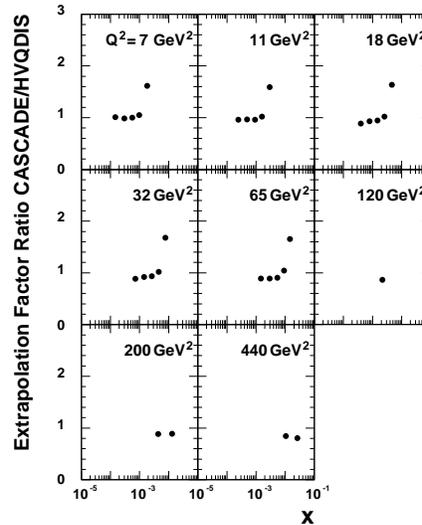}}
 \caption{\label{fig:extraPol} The extrapolation factors from CASCADE are normalised to the ones obtained by HVQDIS.}    
\end{wrapfigure}
production to the visible $D^{*\pm}$ meson cross sections is estimated to be of the order of \mbox{1-2\%} and is therefore neglected. The extraction of $F^{c~\rm{exp}}_2$ is faced with an intrinsic problem as the measurement covers about 30\% of the total phase space for charm production. The ratio $\sigma_{\rm{full}}^{\rm{theo}} / \sigma_{\rm{vis}}^{\rm{theo}}$ is the extrapolation factor to the full phase space which depends strongly on the underlying model. The ratio of the extrapolation factors from CASCADE normalised to HVQDIS is shown in Figure \ref{fig:extraPol}, where in general an agreement between the models at the level of 10\% is observed. This is not the case for the highest $x$ region, where differences in extrapolation of 80\% are observed, which is known to depend strongly on the $\eta$ region of the $D^*$ cross section measurement. If extended to $|\eta(D^*)| < 1.9$ this differences go down to 20\%~\cite{dis09:jung2}.

\subsection{Extrapolation Uncertainties}
The extrapolation needed to determine $F^{c~\rm{exp}}_2$ depends significantly on the underlying model and introduce extrapolation uncertainties. These are estimated by varying the model parameters like charm mass, renormalisation and factorisation scales, PDF set and fragmentation parameters. The parameter variations together with the average relative uncertainties are summarised in Table \ref{tab:ParVar}. 
\begin{small}
\begin{table}[htdp]
\begin{center}
{\small
\begin{tabular}{lllll}
\toprule 
name & variation (HVQDIS) & rel. unc.					& variation (CASCADE) & rel. unc. \\ \midrule
charm mass & $1.3 < m_c < 1.6 ~\mathrm{GeV}$ &6\%				& $1.3 < m_c < 1.6 ~\mathrm{GeV}$ & 7\% \\
scale $\mu_{0}^{2} = Q^2 + 4m_{c}^{2}$  & $0.5 < \mu_{r,f}/\mu_{0} <2$ & 4\%	& $0.5 < \mu_{r,f}/\mu_{0} <2$ & 2\% \\
PDF					& CTEQ5f3 \& MRST04ff3& 13\%		& A0$-$, A0 , A0$+$ & 2\% \\
$\hat{s}$ fragmentation:		& 	&			& 				&	\\ 
\cmidrule(r){1-1}
low  $\hat{s} (< 70~\rm{GeV^2})$	& $\alpha = 6.0^{+1.0}_{-0.8}$	&	& $\alpha = 8.2\pm 1.1$ &\\
high $\hat{s} (> 70~\rm{GeV^2})$	& $\alpha = 3.3\pm 0.4$		&5 \%	& $\alpha = 4.6\pm 0.6$ & 8\% \\
$\hat{s}$ threshold			& $70 \pm 20~\rm{GeV^2}$	&2 \%	& $70 \pm 20~\rm{GeV^2}$ & 2\% \\
\bottomrule 
\end{tabular}
}
\end{center}
\caption{\label{tab:ParVar} Uncertainty of quantities which is utilised for the estimation of the extrapolation uncertainty calculated from HVQDIS and CASCADE. The relative uncertainties on the extrapolation factor are summarised.}
\end{table}%
\end{small}

\subsection{Results}
$F^{c}_2$ determined from the inclusive $D^{*\pm}$ cross sections as a function of $x$ for different values of 
\begin{figure}[ht]
   \centerline{\includegraphics[width=0.95\columnwidth]{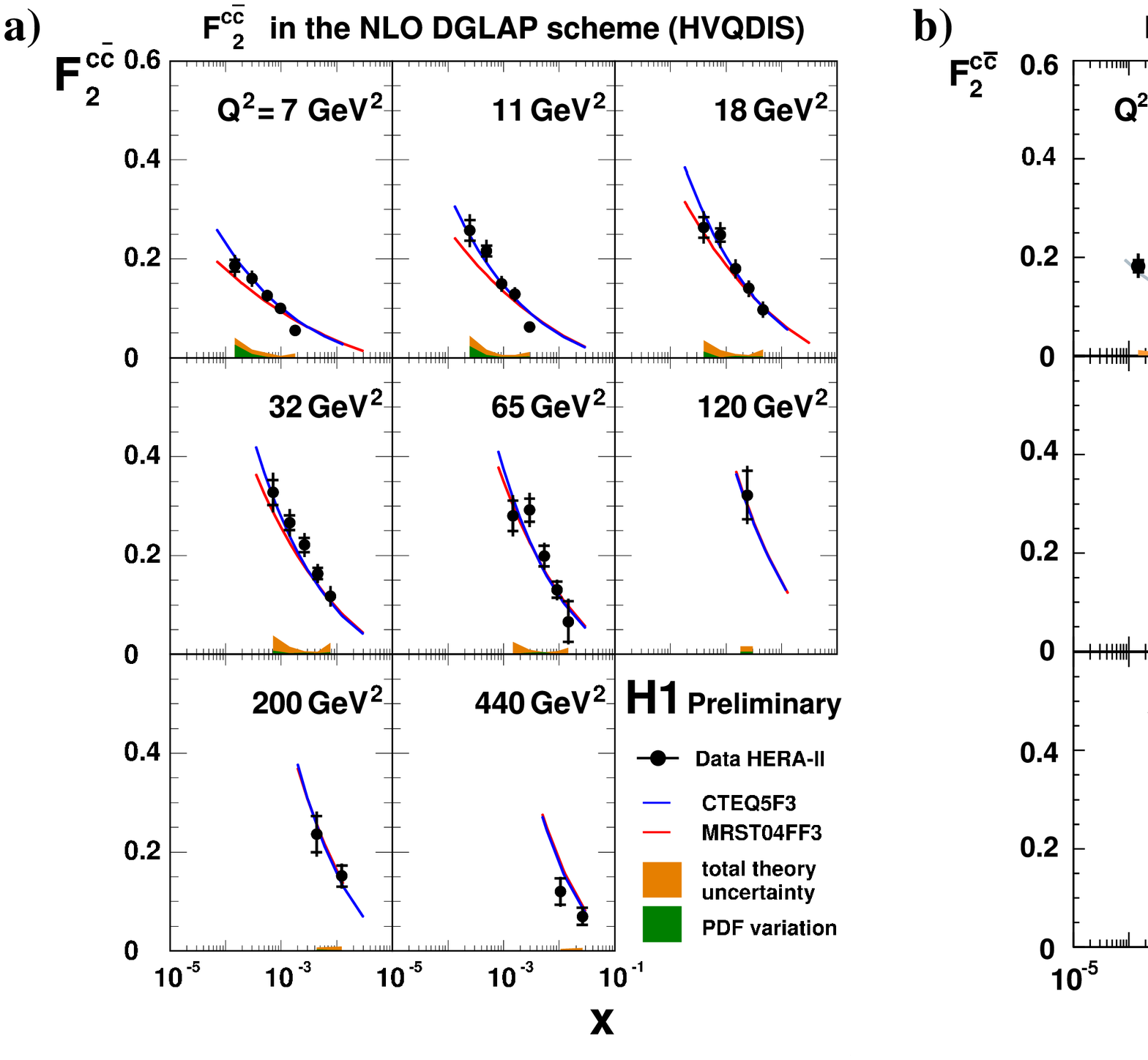}}
 \caption{\label{fig:f2c_ccfm_dglap} $F_2^c$ as determined from the inclusive $D^{*\pm}$ meson cross sections using HVQDIS a) or CASCADE b). The error bars refer to the statistical (inner) and systematical error (outer) added in quadrature. The solid lines represent the predictions of $F_2^c$ using appropriate parton densities for HVQDIS and CASCADE. The shaded band shows the relative change of the central value due to variation of the model parameters.}    
\end{figure}
$Q^2$ is shown in Fig.~\ref{fig:f2c_ccfm_dglap}a) or b) using either HVQDIS or CASCADE for the extrapolation to the full phase space. The inner error bar corresponds to the statistical error, whereas the outer is the systematic error added in quadrature. The relative uncertainties from the fragmentation model are added in quadrature to the systematic uncertainty of the data, whereas all other extrapolation uncertainties are indicated by a shaded band at the bottom of a certain $Q^2$ bin of $F^{c}_2$ as shown in figure \ref{fig:f2c_ccfm_dglap}a) and b). The prediction of HVQDIS using the proton PDF CTEQ5f3 is shown in Fig.~\ref{fig:f2c_ccfm_dglap}a) by the dark solid line and describes the data rather well. In addition also MRST04FF3 was used with HVQDIS, which underestimates the data slightly at low $x$ and $Q^2$. In case of CASCADE the unintegrated gluon distribution A0 was used and is indicated in Fig.~\ref{fig:f2c_ccfm_dglap}b) by the solid line, which describes the data nicely. A comparison of $F^{c}_2$ using HVQDIS or CASCADE shows differences dominantly at high $x$ and especially a steeper slope towards low $x$ for the values extracted with HVQDIS. 

\section{Conclusions}
% By extrapolating the visible $D^{*\pm}$ meson production cross section to the full phase space in $p_{T}(D^*)$ and $\eta(D^*)$, $F^{c}_2$ has been extracted in the framework of pQCD at NLO using HVQDIS in the DGLAP model. $F^{c}_2$ has also been extracted in the CCFM scheme using the Monte-Carlo program CASCADE. For both extractions the extrapolation uncertainty has been estimated by the variation of the input parameters (see \ref{tab:ParVar}) and is sizeable compared to the statistical and systematic error of the cross section data. Especially the uncertainty of the fragmentation model has been estimated from results of an H1 measurement.\\
% The comparison of the data to the two models shows good agreement. The difference between the $F^{c}_2$ extracted in the two schemes is large compared to the uncertainties within each of the models showing the need to measure a larger part of the phase space in order to reduce these model uncertainties.

The charm contribution, $F^{c}_2$, to the proton structure was determined using the measured cross sections of $D^*$ production in $y$ and $Q^2$. Two theoretical models, HVQDIS and CASCADE, were used for the extrapolation in $p_{T}(D^*)$ and $\eta(D^*)$ of the visible cross section to the full phase space. Differences between the extrapolation factors of these models are observed at high $x$. The uncertainty of the fragmentation model has been estimated from results of an H1 measurement. The total theory uncertainties are much larger than the experimental errors. Although especially at low $x$ the predictions from HVQDIS with different proton PDFs differ slightly the discriminating power of the current analysis is not sufficient. Otherwise the comparison of the data to the predictions of HVQDIS and CASCADE shows a good agreement. 

\begin{footnotesize}
% IF YOU DO NOT USE BIBTEX, USE THE FOLLOWING SAMPLE SCHEME FOR THE REFERENCES
% ----------------------------------------------------------------------------
% ****************************************************************************
% BIBLIOGRAPHY AREA
% ****************************************************************************

\bibliographystyle{unsrt}
\bibliography{Jung_AndreasW}

\end{footnotesize}

% ****************************************************************************
% END OF BIBLIOGRAPHY AREA
% ****************************************************************************

\end{document}